\documentclass{sig-alternate}
\usepackage[latin9]{inputenc}
\usepackage[active]{srcltx}
\usepackage{verbatim}
\usepackage{float}
\let\proof\relax

\usepackage{amsmath}
\usepackage{thmtools, thm-restate}
\usepackage{amsthm}
\usepackage{amssymb}
\usepackage{graphicx,kantlipsum,setspace}
\usepackage[belowskip=-15pt,aboveskip=0pt]{caption}
\floatstyle{ruled}
\newfloat{algorithm}{tbp}{loa}
\providecommand{\algorithmname}{Algorithm}
\floatname{algorithm}{\protect\algorithmname}

\makeatletter
\newenvironment{proof*}[1][\proofname]{\par
  \pushQED{\qed}%
  \normalfont \partopsep=\z@skip \topsep=\z@skip
  \trivlist
  \item[\hskip\labelsep
        \itshape
    #1\@addpunct{.}]\ignorespaces
}{%
  \popQED\endtrivlist\@endpefalse
}
\makeatother

\numberwithin{equation}{section}
\numberwithin{figure}{section}
\theoremstyle{plain}
\newtheorem{thm}{\protect\theoremname}
\theoremstyle{definition}
\newtheorem{defn}[thm]{\protect\definitionname}
\theoremstyle{plain}
\ifx\proof\undefined
\newenvironment{proof}[1][\protect\proofname]{\par
\normalfont\topsep6\p@\@plus6\p@\relax
\trivlist
\itemindent\parindent
\item[\hskip\labelsep\scshape #1]\ignorespaces
}{%
\endtrivlist\@endpefalse
}
\providecommand{\proofname}{Proof}
\fi
\theoremstyle{plain}

\theoremstyle{plain}
\newtheorem{cor}[thm]{\protect\corollaryname}
\theoremstyle{remark}

\usepackage{pgfplots}
\usepackage{hyperref}

\makeatletter
\renewenvironment{proof}[1][\proofname]{\par
  \vspace{-\topsep}
  \pushQED{\qed}%
  \normalfont
  \topsep0pt \partopsep0pt 
  \trivlist
  \item[\hskip\labelsep
        \itshape
    #1\@addpunct{.}]\ignorespaces
}{%
  \popQED\endtrivlist\@endpefalse
  \addvspace{6pt plus 6pt} 
}
\makeatother
\usepackage[algo2e,ruled,vlined]{algorithm2e}
\usepackage{tikz}
\usetikzlibrary{calc,intersections} 
\newcommand\markangle[6]{
  \begin{scope}
    \path[clip] (#1) -- (#2) -- (#3);
    \fill[fill opacity=0.5,name path=circle]
    (#1) circle (#4);
  \end{scope}
  \path[name path=line one] (#1) -- (#2);
  \path[name path=line two] (#1) -- (#3);
  \path[%
  name intersections={of=line one and circle, by={inter one}},
  name intersections={of=line two and circle, by={inter two}}
  ] (inter one) -- (inter two) coordinate[pos=.5] (middle);
  \node at ($(#1)!#5!(middle)$) {#6};
}
\newcommand\wlogg{Without loss of generality }

\makeatother

\providecommand{\corollaryname}{Corollary}
\providecommand{\definitionname}{Definition}
\providecommand{\factname}{Fact}
\providecommand{\remarkname}{Remark}
\providecommand{\theoremname}{Theorem}

\theoremstyle{plain}
\newtheorem{lem}{Lemma}

\makeatletter
\def\@copyrightspace{\relax}
\makeatother

\begin{document}
\setlength{\parskip}{0em}
\setlength{\parindent}{0em}

\makeatother
\global\long\def\tP{\tilde{P}}
\global\long\def\Disc{\mathrm{Disc}}

\global\long\def\norm#1{\left\Vert #1 \right\Vert }

\global\long\def\s#1{#1^{*}}

\global\long\def\onenorm#1{\norm{#1}}

\global\long\def\abs#1{\left| #1 \right|}

\global\long\def\abs#1{\left| #1 \right|}

\global\long\def\d#1#2{#1^{(#2)}}

\global\long\def\f#1#2{#1^{[#2]}}

\global\long\def\Mea{\mathrm{Mea}}

\global\long\def\poly{\mathrm{poly}}

\global\long\def\var{\mathrm{var}}

\global\long\def\sgn{\mathrm{sgn}}

\global\long\def\tO{\tilde{O}}

\global\long\def\tf{\tilde{f}}

\global\long\def\Z{\mathbb{Z}}

\global\long\def\N{\mathbb{N}}

\global\long\def\C{\mathbb{C}}

\global\long\def\R{\mathbb{R}}

\global\long\def\F{\mathbb{F}}

\global\long\def\Q{\mathbb{Q}}

\global\long\def\eps{\varepsilon}

\global\long\def\eqdef{:=}

\global\long\def\ads{O(\frac{1}{n})}

\global\long\def\adb{k^{2}}

\global\long\def\adc{k^{3}}

\title{Efficiently Computing Real Roots of Sparse
Polynomials}

\numberofauthors{2} 

\author{
\alignauthor
Gorav Jindal\\
       \affaddr{Max-Planck-Institut f{\"u}r Informatik}\\
       \affaddr{Germany}\\
       \email{gjindal@mpi-inf.mpg.de}
\alignauthor
Michael Sagraloff\\
       \affaddr{Max-Planck-Institut f{\"u}r Informatik}\\
       \affaddr{Germany}\\
       \email{msagralo@mpi-inf.mpg.de}
}

\maketitle
\begin{abstract}
We propose an efficient algorithm to compute the real roots of a sparse polynomial $f\in\R[x]$ having $k$ non-zero real-valued coefficients. 
It is assumed that arbitrarily good approximations
of the non-zero coefficients are given by means of a coefficient oracle.
For a given positive integer $L$, our algorithm returns disjoint
disks $\Delta_{1},\ldots,\Delta_{s}\subset\C$, with $s<2k$, centered at the real
axis and of radius less than $2^{-L}$ together with positive integers
$\mu_{1},\ldots,\mu_{s}$ such that each disk $\Delta_{i}$ contains
exactly $\mu_{i}$ roots of $f$ counted with multiplicity. In addition,
it is ensured that each real root of $f$ is contained in one of the
disks. If $f$ has only simple
real roots, our algorithm can also be used to isolate all real roots.

The bit complexity of our algorithm is polynomial in
$k$ and $\log n$, and near-linear in $L$ and $\tau$, where $2^{-\tau}$
and $2^{\tau}$ constitute lower and upper bounds on the absolute
values of the non-zero coefficients of $f$, and $n$ is the degree of $f$. For root isolation, 
the bit complexity is polynomial in $k$ and $\log n$,
and near-linear in $\tau$ and $\log\sigma^{-1}$, where $\sigma$ denotes
the separation of the real roots. 
\end{abstract}

\section{Introduction}

\subsection{Problem Definition and Contribution}

In this paper, we study the problem of computing the real roots of
a sparse polynomial
\begin{equation}
f(x)=\sum\nolimits_{i=1}^{k}f_{i}x^{e_{i}}\in\R[x],\label{eq:defsparse}
\end{equation}
where $e_{i}$ are non-negative integers, with $0\le e_{1}<e_{2}<\ldots<e_{k}\leq n$,
and $2^{-\tau}\leq|f_{i}|\leq2^{\tau}$ for all $i.$ We call such
a polynomial $f$ an \emph{$(n,k,\tau)$-nomial} or simply a \emph{$k-$nomial}
if $n$ and $\tau$ are either not specified or clear from the context. We may assume that
$k\ge2$ and $e_{1}=0$ as $1-$nomials do not have any real root
different from $0$ and as $f\cdot x^{-e_{1}}$ has exactly the same
roots as $f$ except for a possible root at $0$. We further assume
that, as input, we receive the exponents $e_{i}$ as well as approximations
$\tilde{f}_{i}$ of the non-zero coefficients $f_{i}.$ More specifically, we assume the existence of 
a coefficient oracle that, for any positive integer $\kappa$, provides dyadic
approximations $\tilde{f}_{i}=\frac{m_{i}}{2^{\kappa+1}},$ with $m_{i}\in\Z$
and $|f_{i}-\tilde{f}_{i}|<2^{-\kappa}$ for all $i$.
We call such an approximation $\tilde{f}=\sum_{i=1}^{k}\tilde{f}_{i}x^{e_{i}}$
an \emph{(absolute) $\kappa-$bit approximations of $f$.} Notice that the numbers $n$
and $k$ are directly part of the input, whereas this is not the case for $\tau$.
However, we may easily compute (i.e. for a cost bounded by $\tilde{O}(k\tau)$) a good 
approximation $\tilde{\tau}\in\Z$ of $\tau$ with $\tau<\tilde{\tau}<\tau+2$
by asking the oracle for an $\kappa$-bit approximations
$\tilde{f}$ of $f$ for $\kappa=1,2,4,\ldots$ until $|\tilde{f}_i|>2^{-\kappa+1}$ for all $i$. Then, 
$\tilde{\tau}:=\max_i \lceil|\log |\tilde{f}_i||\rceil$ fulfills the above inequality.

Within recent years, the problem of \emph{isolating} all (real) roots of
a (square-free) polynomial has attracted a lot of interest in the literature; e.g. consider~\cite{BeckerS0Y15,McNamee-Pan,Sagraloff201646} and the references therein.
The most efficient algorithms \cite{BeckerS0Y15,MSW-rootfinding2013,Pan:alg,Sagraloff201646} for root isolation
achieve running times that are considered to be near-optimal for dense polynomials (i.e. if
$k$ is of comparable size as $n$) $f\in\R[x]$. For polynomials with integer coefficients,
the best known bound on the bit complexity of this problem is of size
$\tilde{O}(n^{2}\tau)$. The additional cost for refining isolating intervals
to a size less than $2^{-\tau},$ and thus for computing $L$-bit approximations
of all real roots, is $\tilde{O}(n\tau)$; e.g. see~\cite{Kerber2015377,MSW-rootfinding2013,Pan:2013,Sagraloff201646}. 
Notice that, for $k-$nomials with integer coefficients, the above
bounds are not polynomial in the size of the sparse input representation
of $\tilde{f}$, which is bounded by $O(k(\log n+\tau))$ as we need
$\log n$ bits to store each exponent $e_{i}$ and $\tau+1$ bits
to store each $f_{i}$.
Hence, it is natural to ask whether there exists an
algorithm for either root isolation or approximation that runs
in polynomial time in the size of the sparse input representation.
In \cite{CUCKER1999}, Cucker et al. showed how to compute all integer
roots of a sparse integer polynomial in polynomial time. Lenstra \cite{lenstra99}
further improves upon this result giving a polynomial time
algorithm to compute all rational factors of $f$ of a fixed constant degree.
Furthermore, for polynomials with only a very few
non-zero coefficients, there exist polynomial time algorithms to approximate
(and also count) the real roots of $f.$ Rojas and Ye \cite{Rojas05,Ye1994271}
propose an algorithm for $3$-nomials that uses only $O(\log n)$
arithmetic operations in the field over $\Q$ generated by the coefficients
of $f$. Bastani et al. \cite{Bastani11} propose a polynomial time
algorithm to count the number of real roots for most $4-$nomials. 

For isolating the roots of a sparse integer polynomial, we recently
proposed a method \cite{Sagraloff14} that has polynomial arithmetic
complexity and whose bit complexity is $\tilde{\Omega}(n\tau\cdot k^{4}).$
The latter bound is also near-optimal for small $k$ as there exists
a family of Mignotte-like $4-$nomials, for which the output
complexity is always lower bounded by $O(n\tau).$ This result already
rules out the existence of a polynomial time algorithm for isolating
the roots of a sparse polynomial, however, it remains an open question
whether counting the real roots or computing $L-$bit approximations
of the real roots can be achieved in polynomial time. 

In this paper, we give a positive answer for a slight relaxation of the latter problem.
That is, we give a polynomial time algorithm to compute a partial clustering of the roots that contains all real roots of $f$.
For a more precise statement, we need the following 
definitions, where $\Delta_{r}(m)\subset\C$
denotes the open disk in complex space with center $m$ and radius
$r$.

\begin{defn}
[$(L,I)$-covering] For a polynomial $f$ as in (\ref{eq:defsparse}), an integer $L\in\N$, and an interval $I\subset\R$, we call a list $((\Delta_{r_{1}}(m_{1}),\mu_{1}),(\Delta_{r_{2}}(m_{2}),\mu_{2}),\ldots,(\Delta_{r_{t}}(m_{t}),\mu_{t}))$
an \emph{$(L,I)$-covering for $f$ }if the following conditions are
fulfilled:\vspace{-0.1cm}
\begin{enumerate}
\item The disks $\Delta_{r_{i}}(m_{i})$
are pairwise disjoint, $m_{j}$ are real values with $m_{1}<\cdots<m_{t}$,
and $r_{j}\leq2^{-L}$ for all $j$.\vspace{-0.1cm}
\item $\Delta_{r_{j}}(m_{j})$ contains exactly $\mu_{j}$ roots of $f$ for all $j$.\vspace{-0.1cm}
\item For every real root $\xi$ of $f$ in $I,$ there exists some disk
$\Delta_{r_{j}}(m_{j})$ that contains $\xi$. 
\end{enumerate}
\end{defn}
We further introduce a weaker version of $L$-covering:
\begin{defn}
[Weak $(L,I)$-covering]A \emph{weak $(L,I)$-covering} for $f$ is
a list $(I_{1},\ldots,,I_{t})$ of open disjoint and sorted real intervals that fulfills
the following conditions:\vspace{-0.1cm}
\begin{enumerate}
\item The width of each interval $I_{j}$ is at most $2^{\ensuremath{-L}}.$ \vspace{-0.1cm}
\item For every real root $\xi$ of $f$ in $I$, there exists an interval
$I_{j}$ that contains $\xi.$ 
\end{enumerate}
\end{defn}

If $I=\R,$ we omit $I$ and just call a (weak) \emph{$(L,\R)$-covering
for $f$ }a (weak) \emph{ $L$-covering for f.}
Then, our main contribution is a polynomial-time algorithm for computing an $L-$covering: 

\begin{thm}\label{thm:main}
For an $(n,k,\tau)$-nomial, we can
compute an $L$-covering $\mathcal{L}$ of size $|\mathcal{L}|< 2k$ in time $\tilde{O}(\poly(k,\log n)\cdot(\tau+L))$.
\end{thm}

Notice that our algorithm computes $L-$bit approximations 
of all real roots but might also return (real-valued) $L-$bit approximations
of some non-real roots with a small imaginary part.
Further notice that unless $\mu_{j}$ is odd, we also do not know whether
$m_{j}$ actually approximates a real root, and unless $\mu_{j}=1,$ we cannot conclude that a disk $\Delta_{r_{j}}(m_{j})$
in an $L$-covering is isolating for a root of
$f$. Hence, in general, our algorithm does not yield the correct
number of distinct real roots. 
However, if $f$ has only simple roots, we may compute an $L-$covering
for $f$ for $L=2,4,8,\ldots$ until $\mu_{j}=1$ for all $j$. Then,
the disks $\Delta_{r_{i}}(m_{i})$ isolate all real
roots.

\begin{thm}\label{thm:isolation}
Let $f$ be an $(n,k,\tau)$-nomial with only simple real roots, and let $\sigma$ be the minimal distance between any two (complex) distinct roots of $f$ (i.e. the \emph{separation} of $f$). Then, we can compute isolating intervals for all real roots in $\tilde{O}(\poly(k,\log n)(\tau+\log\max(1,1/\sigma)))$ bit operations.
\end{thm}

We improve upon~\cite{Sagraloff14} in several ways. Namely,~\cite{Sagraloff14} only applies to integer polynomials, whereas
our novel approach applies to polynomials with arbitrary real coefficients.
In addition, the running time of the algorithm in~\cite{Sagraloff14} 
does not adapt to the actual hardness of the roots, whereas the complexity
of our novel approach rather depends on the actual separation than on the worst-case bound \cite{Yap:book}
of size $2^{-O(n(\tau+\log n))}$ for the separation of an integer
polynomial. In the worst case, our method isolates all real roots
of a very sparse integer polynomial (i.e. $k=(\log(n\tau))^{O(1)}$)
in time $\tilde{O}(n\tau)$, and is thus near optimal.;
see~\cite{Sagraloff14}\smallskip

\subsection{Overview of the Algorithm}\label{sec:overview}

Before we go into detail, we give a
brief overview of our algorithm, where we omit technical details. We first remark that the problem
of computing an $(L,[1,\infty))$-covering can be reduced
to the problem of computing an $(L,[0,1])$-covering (in fact, we are computing an $(L,[0,1+1/n])$-covering but this for technical reasons only) by means of the
coordinate transformation $x\mapsto\frac{1}{x}$ followed by multiplication
with $x^{n}$. We may also reduce the problem of computing
an $(L,(-\infty,0])$-covering of $f$ to the problem of computing
an $(L,[0,\infty))$-covering by means of the coordinate transformation
$x\mapsto-x$. Hence, we are eventually left with merging $(L,[0,1])$-coverings
for the polynomials $f,$ $x^{n}\cdot f(1/x),$ $f(-x),$ and $x^{n}\cdot f(-1/x)$
in a suitable manner. We give details for this step 
in Section~\ref{sec:covering}. Notice that the considered coordinate
transformation preserves the sparseness of the input polynomial, hence
we may concentrate on the problem of computing an $(L,[0,1])$-covering
for $f$ only. For this, we first compute a weak $(L,[0,1])$-covering
of $f$, which is achieved by recursively computing weak $(L,[0,1])$-coverings
of the so-called \emph{fractional derivatives of $f$. }
\begin{defn}
[Fractional Derivatives]Let $f$ be a polynomial as in (\ref{eq:defsparse}).
Then, we define $\f f1\eqdef\frac{f^{\prime}}{x^{e_{2}-1}}$ as the
\emph{(first) fractional derivative of $f$.} In other words, we divide
the first derivative $f'$ of $f$ by the highest possible power of
$x$ that divides $f'$. The \emph{$i-$th fractional derivative $f^{[i]}$
of $f$} is then recursively defined as the first fractional derivative
of $f^{[i-1]}.$ Notice that, for $i\le k-1$, $f^{[i]}$ is an $(n,k-i,\tau+k\cdot\log n)-$nomial with a non-zero constant term 
and $f^{[i]}\equiv0$ for $i\ge k.$ We further use the notation $\mathcal{D}_{f}$
to denote the tuple of all non-zero fractional derivatives $f,\f f1,\f f2,\f f2,\ldots,\f f{k-1}$,
i.e, $\mathcal{D}_{f}=(f,\f f1,\f f2,\f f3,\ldots,\f f{k-1})$.
\end{defn}

The general idea of recursively computing the real roots of $f$ from
the real roots of its fractional derivatives has already
been considered in previous work; e.g.~\cite{DBLP:conf/issac/GarciaG12,Collins:1976,Coste2005479,lenstra99,Pan:2007,Rojas05,Sagraloff14}.
The simple idea is that, given a weak $(L,[0,1])$-covering $(I_{1}',\ldots,I_{t'}')$
for $f^{[1]}$, we already know that in between two consecutive intervals
$I_{j}=(a,b)$ and $I_{j+1}=(c,d)$, the polynomial $f$ is monotone,
and thus there can be at most one real root in between $b$ and $c$, which then must be simple.
In order to check for the existence of such a root, it suffices to
check whether $f$ changes signs at the points $b$ and $c.$ In case
of a sign change, we may then refine the interval $(a,b)$, which
is known to be isolating for a real root of $f$, to a width less
than $2^{-L}.$ If we proceed in this way for all intervals
in between two consecutive intervals as well as with the leftmost
interval, whose endpoints\footnote{For technical reasons, we will indeed consider slight perturbations of $0$ and $1$ in our algorithm.} are $0$ and the left endpoint of $I_{1}'$,
and the rightmost interval, whose endpoints are the right endpoint
of $I_{t'}'$ and $1,$ then we obtain a set of intervals $I_{j}''$
of size at most $2^{-L}$ that cover all real roots of $f$ that are
contained in $[0,1]$ but in none of the intervals $I_{j}'$. Hence,
the union of the intervals $I_{j}'$ and $I_{j}''$ constitutes an
$(L,[0,1])$-covering for $f$. This shows how to compute an $(L,[0,1])$-covering
for $f$ from recursively computing $(L,[0,1])$-coverings for its
fractional derivatives. 

We remark that, in this simplistic description, we have omitted several
key problems one faces when formalizing the algorithm: Evaluating the sign 
of a polynomial $f$ at given points $b,c$ may
require a very high precision, which should be avoided to ensure a polynomial bit complexity. In addition, we need an efficient
refinement method that uses only a polynomial number of iterations.
For the latter problem, we use a slightly modified variant of our algorithm
from \cite{Sagraloff14,Sagraloff201646}. For the computation of the
sign of $f$ (and its higher order fractional derivatives) at certain
points, we consider an approach that allows us to slightly perturb
the evaluation points such that the absolute value of each of the
considered polynomials does not become too small. One major contribution
of this paper, when compared to our previous work~\cite{Sagraloff14},
is to show that this can be done in way such that the precision
always stays polynomial in $\log n,$ $k,$ $\tau,$ and $L$. 

In the second step, we derive an $(L,[0,1])-$covering
from a weak $(L',[0,1])-$covering, where $L'$ has been chosen sufficiently
large. A straight forward approach would be to use a
method for computing the number of roots in the \emph{one-circle region
$\Delta(I)=\Delta_{r}(m)$} of each interval $I$ in the weak $(L',[0,1])-$covering.
Here, $\Delta(I)$ is defined as the disk centered at the midpoint
$m=m(I)$ of $I$ and passing through the endpoints of the interval.
In the literature, several methods have been proposed to count the number
of roots in a disk in complex space. Unfortunately, these algorithm
are not sparsity aware, which rules out a straight-forward application
of them. Recent work \cite{BeckerS0Y15} introduces the so-called $T_l$-test, a method for root counting 
based on Pellet's Theorem. The method only needs
to compute approximations of the coefficients of the polynomial $f(m+r\cdot x)$, however, we cannot afford to compute all coefficients. Fortunately, in our situation, only the first $k^2$ coefficients are actually needed to determine the outcome of the test. In
order to guarantee success of the test, it may further be necessary
to merge some of the intervals in the weak covering and to consider
disks that are larger than the one-circle regions of the merged intervals.
This explains why we need a weak $(L',[0,1])-$covering with
a sufficiently large $L'>L.$ We consider our method for counting the roots
of a sparse polynomial in a disk as the second
main contribution of our paper. \vspace{-0.1cm}

\section{On the Geometry of Roots}\label{sec:geometry}

Descartes' Rule of Signs states that the number $\var(F)$ of sign changes in
the coefficient sequence of a polynomial $F\in\R[x]$ constitutes an upper bound
on the number of real roots (counted with multiplicity). Hence, it
follows immediately that a $k-$nomial $f$ as in (\ref{eq:defsparse}) has at most $k-1$ negative
and at most $k-1$ positive real roots. Apart from this simple fact,
$k$-nomials have indeed much more structure on their roots, which
we will briefly survey in this section. 

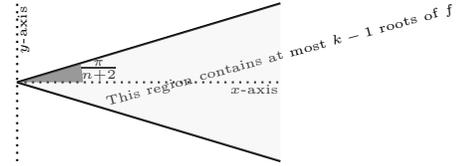
\begin{figure}[t]
\centering
\begin{centering}
\begin{tikzpicture}[scale=0.35, thick]
\coordinate (O) at (0,3);
\coordinate (O1) at (0,0);
\coordinate (O2) at (0,6);
\coordinate (A) at (10,6); 
\coordinate (B) at (10,0); 
\coordinate (C) at (10,3);
\draw (O)--(A);
\draw (O)--(B);
\draw [dotted](O)--(C);
\draw [dotted](O1)--(O2);
\markangle{O}{A}{C}{25mm}{30mm}{\tiny \text{ }$\frac{\pi}{n+2}$}
\node[rotate=15] at (10,4) {\tiny This region contains at most $k-1$ roots of $f$};
\node at (9,2.75) {\tiny $x$-axis};
\node [rotate=90]at (0.25,5) {\tiny $y$-axis};
 \fill[gray!20,nearly transparent]  (B) -- (O) -- (A) -- cycle;
\end{tikzpicture}
\par\end{centering}\vspace{0.25cm}
\centering{}\caption{\label{fig:Cone-1}{\small 
The cone $C_n$ contains at most $k-1$ roots of $f$.
}}
\end{figure}

Let $I=(a,b)$ be an interval, $F_{I}(x):=(x+1)^{n}\cdot F\left(\frac{ax+b}{x+1}\right)$, and
$v_{I}\eqdef\var(F,I)$ be the number of sign changes in the coefficient sequence
of the polynomial $F_{I}$.
Notice that there is a one-to-one correspondence between the roots
of $F$ in $I$ and the positive real roots of $F_{I}$ via the M{\"o}bius
transformation that maps a point $x\in\C\setminus\{-1\}$ to $\frac{ax+b}{x+1}\in\C$.
Thus, $v_{I}$ constitutes an upper bound on the number of roots of
$F$ in $I.$ 
In fact, $v_I$ also constitutes a lower bound on the number of roots in the so called  \emph{Obreshkoff lens} $L_n$ of the interval $I$.
$L_n$ is defined as the intersection 
$L_n:=\overline{C}_{n}\cap \underline{C}_{n}$ of the two open disks $\overline{C}_{n},\underline{C}_{n}\subset \mathbb{C}$ that intersect the real axis in 
the endpoints $a$ and $b$ of $I$, and whose centers see
the line segment $(a,b)$ under the angle $\frac{2\pi}{n+2}$. For an illustration, see~\cite[Fig.~1]{Sagraloff201646}.
It further holds~\cite{Sagraloff14,Sagraloff201646}) that $\var(F,I)\le \var(F)\le k-1$ for any interval $I\subset\mathbb{R}^+$, hence we conclude that the Obreshkoff lens $L_n$ of any such interval contains at most $k-1$ roots. For $b\mapsto\infty$, the Obreshkoff lens $L_{n}$ of the interval
$I=(0,b)$ converges to the cone $C_{n}$ whose boundary are the two
half-lines starting at the origin and intersecting the real axis at
an angle $\pm\frac{\pi}{n+2}$; see Figure~\ref{fig:Cone-1}. Hence, it follows that the interior of $C_{n}$ contains
at most $k-1$ roots of any given $k-$nomial of degree $n$. 

\begin{thm}
\label{lem:Cone--contains}The cone $C_{n}$ contains at most $k-1$ roots
of any $k$-sparse polynomial of degree $n$.
\end{thm}

\section{Polynomial arithmetic}\label{sec:arithmetic}

Our algorithm only needs to perform basic operations on polynomials.
In particular, we need to evaluate the sign of a given sparse polynomial
at some points $x$. As we already mentioned in the overview of our algorithm,
the complexity of this operation becomes too large if the value of
the polynomial at a given point $x$ is almost zero as then one needs to
perform computations with a very high precision. Also,
exact evaluation of a sparse polynomial at a rational point (even
of small bitsize) is expensive as the output has bitsize
linear in $n$. Instead, we consider approximate evaluation, which
allows us to evaluate a sparse polynomial $f$ as in (\ref{eq:defsparse})
at an arbitrary point $x\in(0,1+1/n)$ to an absolute error less than
$2^{-L}$ in a time that is polynomial\footnote{Notice that, for $c\in(0,1+1/n^{O(1)})$, we may omit the term $n\log\max(1,|c|)$ in the bounds stated in Lemma~\ref{lem:Let--be}.} in $\log n,$ $k,$ $\tau,$
and $L.$ More precisely, we derive the following result:
\begin{restatable}{lem}{polyar}
\label{lem:Let--be}Let $f\in\R[x]$ be an $(n,k,\tau)$-nomial, $c$
be a positive real number, and $L$ a non-negative integer. Then,
we can compute an $L$-bit approximation $\lambda$ of $f(c)$ (i.e.
$|\lambda-f(c)|<2^{-L}$) in a number of bit operations bounded by
\[
\tilde{O}((k+\log n)\cdot(L+n\log\max(1,|c|)+\log n+\tau+k)).
\]
\end{restatable}

\begin{proof*}
In essence, we follow the same approach as in \cite{Kerber2015377}.
That is, for a fixed non-negative integer $K$, we perform each occurring
operation $\circ$ (i.e. either addition or multiplication) with fixed
precision $K$. More precisely, the input is initially rounded after
the $K$-th bit after the binary point. Then, in each of the following
steps, we replace each exact operation $\circ$ between two numbers
$a$ and $b$ by a corresponding approximate operation $\tilde{\circ}$,
where we define $a\tilde{\circ}b$ to be the result obtained by rounding
$a\circ b$ after the $K$-th bit after the binary point. Suppose
that we have computed approximations $\tilde{a}=a+\eps_{1}$ and $\tilde{b}=b+\eps_{2}$
of two intermediate results $a$ and $b$, where we assume that $\eps:=\max(|\eps_{1}|,|\eps_{2}|,2^{-K})<1.$
Then, we have $|a\cdot b-\tilde{a}\tilde{\cdot}\tilde{b}|<|a|\cdot|\eps_{1}|+|b|\cdot\eps_{2}+|\eps_{1}\eps_{2}|+2^{-K}<4\cdot\eps\cdot\max(1,|a|,|b|)$
and $|a+b-\tilde{a}\tilde{+}\tilde{b}|<|\eps_{1}|+|\eps_{2}|+2^{-K}$.
Hence, when evaluating one term $f_{i}\cdot x^{e_{i}}$ of $f$ at
the point $x=c$ with absolute precision $K>L+\log k+1+\tau+(2\log n+1)\cdot(n\cdot\log\max(1,|c|)+2)$
via repeated squaring, we induce a total error $\eps_{i}$ for the
computation of $f_{i}\cdot c^{e_{i}}$ of size less than $2^{\tau}c^{n(2\log n+1)}\cdot4^{2\log n+1}\cdot2^{-K-L}<(2k)^{-1}\cdot2^{-L}$
as there are at most $2\log n+1$ multiplications, and each (exact)
intermediate result has absolute value bounded by $\max(2^{\tau},c^{n}).$
When eventually summing up the approximations of all terms $f_{i}\cdot x^{e_{i}}$,
we thus induce an error of size less than $\sum_{i}\eps_{i}+k\cdot2^{-K}<2^{-L}$
for the computation of the final result. The bound on the bit complexity
follows from the fact that we need $O(k+\log n)$ arithmetic operations
on integers of bitsize $O(K+\tau+\log k+n\log\max(1,|c|))=O(K)$,
and each such operation uses $\tilde{O}(K)$ bit operations.

\end{proof*}

We already mentioned that evaluating the sign of a polynomial $f$
at a point $x$ might be costly if $f(x)$ has a small absolute value.
In order to avoid such undesired computations, we first
perturb $x$ in a suitable manner. That is, instead
of evaluating the sign of $f$ at $x,$ we evaluate its sign at a
nearby point, where $f$ becomes large enough. This can be done in a way such that the actual behavior of the algorithm does not change. We will call such points ``admissible''. We remark that we already used this
concept in previous work \cite{Sagraloff14,Sagraloff201646}. Here,
we modify the approach to choose an admissible point, where the sign
of each fractional derivative of a sparse polynomial $f$ can be evaluated
in polynomial time.
\begin{defn}
[Admissible point]Let $g:\R\rightarrow\R$ be a function and $m[t;\delta]=\{m_{i}\eqdef m+(i-k)\cdot\delta;i=0,1,\ldots,2t\}$
be a multipoint. We call a point $\s m\in m[t;\delta]$ to be \emph{$(g,m[t;\delta])$-admissible}
if $\abs{g(\s m)}\geq\frac{1}{8}\cdot\max_{x\in m[t;\delta]}\abs{g(x)}$.
\end{defn}
If $t$ and $\delta$ (or even $m$ and $g$) are clear from the context, we simply call a $(g,m[t;\delta])$-admissible point
$(g,m)$-admissible (or just admissible). Since the value of a polynomial $g$ at an admissible
points is ``relatively large'', we expect that $g$ has no root in a neighborhood of an admissible point. The
following lemma formalizes this intuition. 
\begin{restatable}{lem}{norootinsmalldisk}
\label{lem:If--to}Suppose that $m\in\R_{+}$ and $\s m\in m[t;\delta]$
is an $(f,m[t;\delta])$-admissible point for an $(n,k,\tau)$-nomial
$f$ with $k\ge2$ and $k\le t\leq k^{2}$. Further assume that $\frac{m}{\delta}>4k^{2}n^{2}$. Then, the disk \textup{$\Delta_{\frac{\delta}{k^{4k}}}(\s m)$ }does
not contain any root of $f$. 
\end{restatable}

\begin{proof*}
Let $z_{1},\ldots,,z_{n}$ denote the complex roots of $f.$ Since
$f(x)$ has at most $k-1$ roots in the cone $C_{n}$ (see Figure~\ref{fig:Cone-1}
) and $t\ge k$, there exists a point $b\in m[t;\delta]$ such that
$\Delta_{\delta/2}(b)$ does not contains any of these $k-1$ roots.

By way of contradiction assume that there is a root $z_{l}$ of $f$
in the ball of radius $\frac{\delta}{k^{4k}}$ around $\s m$.

We have

\[
\frac{f(\s m)}{f(b)}=\prod_{i=1}^{n}\frac{\s m-z_{i}}{b-z_{i}}
\]

By using the triangle inequality for the distance between $\s m$
and $z_{i}$, one can see that, for the roots $z_{i}\neq z_{l}$ that
are contained in $C_{n}$, we have 
\[
\frac{\s m-z_{i}}{b-z_{i}}\leq2t+1\leq2k^{2}+1,
\]

whereas $\frac{m^{*}-z_{l}}{b-z_{l}}<2\cdot k^{-4k}.$ Now, consider
the roots $z_{i}$ of $f$ that are outside of $C_{n}$. Since $\frac{m}{\delta}>4k^{2}n^{2}$
, it follows that the distance of $\s m$ to any such $b$ is at at
least $32k^{2}\delta n$.

Again using the triangle inequality for distance between $\s m$ and
$z_{i}$, this implies that 

\[
\frac{\s m-z_{i}}{b-z_{i}}\leq1+\frac{2t}{2k^{2}n}\leq1+\frac{1}{n}
\]

Hence

\begin{eqnarray*}
\frac{f(\s m)}{f(b)} & \leq & \left(1+\frac{1}{n}\right)^{n-k+1}\cdot(2k^{2}+1)^{k-2}\cdot\frac{2}{k^{4k}}\\
 & < & \left(1+\frac{1}{n}\right)^{n-k+1}\cdot(\frac{2k^{2}+1}{k^{4}})^{k-2}\cdot\frac{2}{k^{8}}\\
 & \leq & \frac{2e}{k^{8}} <  \frac{1}{8}.\mbox{ }
\end{eqnarray*}

This contradicts the fact that $\s m$ is an admissible point.
\end{proof*}

\begin{defn}
\label{def:Let--be}Let $\mathcal{G}=(g_{1},g_{2},\ldots,g_{t})$
be a tuple of $t$ functions $g_{i}:\R\rightarrow\R$. Then, $M_{\mathcal{G}}(x)$
is defined as follows:
\[
M_{\mathcal{G}}(x)\eqdef\min(\abs{g_{1}(x)},\abs{g_{2}(x)},,\ldots,\abs{g_{t}(x)}).
\]
For a fixed real $x$,
we call $\mathcal{\tilde{G}}(x)=(\tilde{g}_{1}(x),\tilde{g}_{2}(x),\ldots\tilde{g}_{t}(x))$
an \emph{$L$-approximation of $\mathcal{G}(x)$} if $\abs{\tilde{g}_{i}(x)-g_{i}(x)}\leq2^{-L}$
for all $i$.
\end{defn}
We first show how to compute an admissible point $m^{*}\in m[t;\delta]$
for $M_{\mathcal{G}}(x)$ under the assumption that we
can compute an $L$-approximation of $\mathcal{G}(x)$ for any $x\in m[t;\delta]$
in time $T(L)$.
\begin{restatable}{lem}{admissiblepointcomp}
\label{lem:Let--be-1}Let $\mathcal{G}=(g_{1},g_{2},\ldots,g_{t})$
be as in Definition \ref{def:Let--be}, $m[t;\delta]$ a multipoint
and $m_i:=\max_{a\in m[t;\delta]}\abs{M_{\mathcal{G}}(a)}$. Suppose that for
for a point $x\in m[t;\delta]$ we can compute an $L$-approximation
of $\mathcal{G}(m_i)$ in time $T(L,x)$, then we can compute an $(M_{\mathcal{G}},m[t;\delta])$-admissible
point $m^{*}\in m[t;\delta]$ in time $$O(t\cdot\log\log\max(\lambda^{-1},1)\cdot(T(\log\max(\lambda^{-1},1))).$$
Within the same time, we may compute an integer $\ell^{*}$ with 
$
2^{\ell^{*}-1}\leq\abs{M_{\mathcal{G}}(m^{*})}\leq\lambda\leq2^{\ell^{*}+1}.
$
\end{restatable}

\begin{proof*}
We proceed in the same fashion as in Lemma 8 of \cite{Sagraloff201646}.
For $L=1,2,4,\ldots,$, we compute $L$-approximations $\tilde{\mathcal{G}}_{i}^{L}=(\tilde{g}_{1}^{L}(m_{i}),\tilde{g}_{2}^{L}(m_{i}),\ldots\tilde{,g}_{t}^{L}(m_{i}))$
of $\mathcal{G}(m_{i})$ for all points $m_{i}\in m[t;\delta]$ until
the following condition is satisfied for at least one $m_{i}$:
\[
M_{i}^{L}:=\min(\tilde{g}_{i}^{L}(m_{i})|,|\tilde{g}_{2}^{L}(m_{i})|,\ldots,|\tilde{g}_{t}^{L}(m_{i})|)\geq4\cdot2^{-L}=2^{-L+2}.
\]
Then, let $i_{0}$ be the index such that $M_{i_{0}}^{L}$ is maximal
among all $M_{i}^{L}$, and let $\ell^{*}$ be an integer such
that $\abs{\ell^{*}-\log M_{i_{0}}^{L}}\leq\frac{1}{2}$. We output
$\ell^{*}$ and $m^{*}:=m_{i_{0}}$ . 
It is now straight-forward (cf. the proof of Lemma~8 in~\cite{Sagraloff201646}) to show that $2^{\ell^{*}-1}\leq M_{\mathcal{\mathcal{G}}}(m^{*})\leq\lambda\leq2^{\ell^{*}+1}$. 

Following the above approach, we must succeed for an $L\leq2\log\max(\frac{1}{\lambda},1)$.
Since we double $L$ at most $O(\log\log\max(\frac{1}{\lambda},1))$
many times and since we approximately evaluate the functions $g_{i}$
at $t$ points, the stated complexity bound follows. 
\end{proof*}

We now apply the above lemma to the special case where
$\mathcal{G}=\mathcal{D}_{f}$ is the sequence of fractional derivatives of
$f$. Then, Lemma \ref{lem:Let--be} yields a bound
of the bit complexity of computing $L$-approximations of $\mathcal{D}_{f}(m_{i})$
for all points $m_{i}\in m[t;\delta]$.
\begin{cor}
\label{cor:Assume-that-}Assume that $f(x)$ is a $(n,k,\tau)$-nomial,
$m[t;\delta]$ a multipoint and $\lambda:=\max_{m_i\in m[t;\delta]}\abs{M_{\mathcal{D}_{f}}(m_i)}$.
Further assume that $m[t;\delta]\subset(0,\alpha)$ for some positive
real number $\alpha$. Then, we can determine an $(M_{\mathcal{D}_{f}},m[t;\delta])$-admissible
point $m^{*}$ and an integer $\ell^{*}$ with 
\[
2^{\ell*-1}\leq\abs{M_{\mathcal{D}_{f}}(m^{*})}\leq\lambda\leq2^{\ell^{*}+1}
\]
using $\tilde{O}(t\cdot k\cdot (k+\log n)\cdot (\tau+k\log n+n\log\max(1,\alpha)+\log\max(1,\lambda^{-1})))$
many bit operations. 
\end{cor}
Notice that the running time of the above algorithm depends on the
value $\lambda:=\max_{m_{i}\in m[t;\delta]}\abs{M_{\mathcal{D}_{f}}(m_{i})}$.
We will now derive a bound on $\lambda$ that shows
that, for a sufficiently large $t$ and suitably chosen $m$ and $\delta$, we can always compute an
$(M_{\mathcal{D}_{f}},m[t;\delta])$-admissible point $m^{*}$ in polynomial
time.
\begin{lem}
\label{lem:Let--be-2}Let $f\in\R[x]$ be a $(n,k,\tau)$-nomial as
in (\ref{eq:defsparse}), and let $a,r$ be positive real numbers
with $r<a$ and such that $(a-r,a+r)$ does not contain any real
root of any fractional derivative of $f(x)$. Then, it holds that
$$|M_{\mathcal{D}_{f}}(a)|=2^{-O(k(k\log n+\tau+\log\max(1,\frac{1}{r})+n\log\max(1,a+r)))}.$$
\end{lem}

\begin{proof*}
We may assume that $r$ is small enough to guarantee that $\frac{a}{r}>2n$. This implies that, for any two points $x,x'\in I_1:=(a-r,a+r)$, we have that $x/x'\in (1-1/n,1+1/n)$. Now, let us write $f=c+x^j\cdot g$ with a constant $c$ of absolute value at least $2^{-\tau}$ and $g$ an $(n-j,k-1,\tau+\log n)-$nomial that is not divisible by $x.$
Then, it holds that $f^{[1]}=j\cdot g+x\cdot g'$, and thus $f'=x^{j-1}\cdot f^{[1]}$. In addition, since $I_1:=(a-r,a+r)$ does not contain any root of $f$ and $f^{[1]}$,
it follows that $f$ is monotone on $I$ and only takes positive or negative values.
This implies that $|f(t)-f(t')|=||f(t)|-|f(t')||$ for all $t,t'\in I$. In addition, for any $t\in I_2:=(a-r/2,a+r/2)$, we can 
choose a point $t'=t\pm r/2$ such that $|f(t)|>|f(t')|$. Now, according to the mean value theorem, there exists a $\xi$ in between
$t$ and $t'$ with $f(t)-f(t')=(t-t')\cdot f'(\xi)=\frac{r}{2}\cdot \xi^{j-1}\cdot f^{[1]}(\xi)$. Hence, we obtain
$|f(t)|>|f(t)|-|f(t')|=||f(t)|-|f(t')||
=|f(t)-f(t')|\ge \frac{r}{2}\cdot \xi^{j-1}\cdot f^{[1]}(\xi)\ge \frac{r}{8}\cdot t^{j-1}\cdot f^{[1]}(\xi),
$
where the latter inequality follows from $(\xi/t)^{j-1}>(1-1/n)^n>1/2$.
Also, $|f(t)|\ge |c|-t^j\cdot |g(t)|\ge 2^{-\tau}-t^{j-1}\cdot k\cdot 2^{\tau+\log n}\cdot \max(1,a+r)^{n}$.
With $\eps:=\min(1,\inf_{x\in I_1} |f^{[1]}(x)|)$, the above inequalities thus imply that 
\begin{align*}
|f(t)|&>\max(\frac{r\eps}{8}\cdot t^{j-1},2^{-\tau}-t^{j-1} k 2^{\tau+\log n}\cdot \max(1,a+r)^{n})
\end{align*}
Now, if $t^{j-1}< 2^{-\tau-1}(k 2^{\tau+\log n}\cdot \max(1,a+r)^{n})^{-1}$, then the second argument in the above term becomes larger than $2^{-\tau-1}$. Otherwise, the first term becomes larger than $\frac{r\eps}{8}\cdot 2^{-\tau-1}(k 2^{\tau+\log n}\cdot \max(1,a+r)^{n})^{-1}$. Hence, we conclude that
\[
\inf_{x\in I_2}|f(x)|>r\cdot\eps\cdot 2^{-2\tau-1-2\log n-n\log\max(1,a+r)}.
\]
We now recursively apply the above result to the fractional derivatives $f^{[k-i]}$ and the intervals $I_{i}:=(a-\frac{r}{2^{i-1}},a+\frac{r}{2^{i-1}})$, where $i=1,2,\ldots,k$. Notice that each of the polynomials is an $(n,k,\tau+k\log n)-$nomial and that $\inf_{x\in I_{1}} |f^{[k-1]}(x)|>2^{-\tau}$ as $f^{[k-1]}$ is a constant of absolute value at least $2^{-\tau}$. Hence, it follows that
\[
\inf_{x\in I_i} |f^{[k-i]}(x)|>2^{-\tau-i\cdot(2\tau-1-2k\log n-n\log\max(1,a+r))}\cdot\prod_{j=1}^{i-1} \frac{r}{2^j}.
\]
\end{proof*}
\medskip
Combining the above lemma and Corollary \ref{cor:Assume-that-} now yields

\begin{thm}\label{thm:Let--be-1}Let $f$ be a $(n,k,\tau)$-nomial as in~(\ref{eq:defsparse}), and let $m[t;\delta]$ be a multipoint with $t\ge k^2$ and $m[t;\delta]\subset(0,\alpha)$ for some for some real number $\alpha$. Then, we can compute an $(M_{\mathcal{D}_{f}},m[t;\delta])$-admissible
point $m^*$ using $\tilde{O}(t\cdot k^2\cdot(k+\log n)\cdot(k\log n+\tau+\log \max(1,\frac{1}{\delta})+n\log\max(1,\alpha)))$ bit operations.
\end{thm}
\begin{proof*}
Since each fractional derivative of $f$
has at most $k-1$ positive real roots and since $t\ge k^{2}$, there exists
an $a\in m[t;\delta]$ such that $(a-\delta/2,a+\delta/2)$ does not contain
any real root of any of fractional derivative. Hence, Lemma \ref{lem:Let--be-2} implies that
$\lambda:=\max_{x\in m[t;\delta]} |M_{\mathcal{D}_f}(x)|\ge  |M_{\mathcal{D}_f}(a)|$ is lower bounded by $2^{-O(k(k\log n+\tau+\log\frac{1}{\delta}+n\log\max(1,a+\delta)))}$.
Corollary \ref{cor:Assume-that-} then yields the claimed
bound on the running time.
\end{proof*}

\section{Refinement}
\label{sec:Refinement}

A crucial subroutine of our overall algorithm is an efficient method for refining an interval $I_0=(a_0,b_0)\subset\R_+$, with $\max(|\log a_0|,|\log b_0|)=O(\tau)$, 
that is known to be isolating for a simple real root of a $k$-nomial $f$. It is assumed that the algorithm receives the sign of $f$ at the endpoints of $I_0$ as additional input.
For the refinement, we consider the algorithm \textsc{NewRefine} from 
Section~3 in~\cite{Sagraloff14} (see also Section~5  in~\cite{Sagraloff201646}), however, we make a single (minor) modification. As the argument from~\cite{Sagraloff14} directly applies, we only state the main results and refer the reader to~\cite{Sagraloff14} for details. 

\textsc{NewRefine} recursively refines $I_0$ to a size less than $2^{-L}$ using a trial and error approach that combines Newton iteration and bisection. For 
this, only $f$ and its first derivative $f'$ need to be evaluated. More precisely, in each iteration, the algorithm computes 
$(f,m[\lceil k/2\rceil;\delta])-$admissible points $m^*$ for a constant number of points $m\in I$ and a corresponding $\delta$ of size $2^{-O(\tau+\log n+L)}$. In addition, $f$ and $f'$ are evaluated at these admissible points to an absolute 
precision that is bounded by $O(\log \max(1,|f(m^*)|^{-1})+\log n+L+\tau)$. Each endpoint of the interval returned by \textsc{NewRefine} is then either one of the admissible points computed in a previous iteration or one of the endpoints of $I_0$.

We now propose the following modification of \textsc{NewRefine}, which we
denote \textsc{NewRefine$^*$}: Whenever \textsc{NewRefine} asks for an
$(f,m[\lceil k/2\rceil;\delta])-$admissible point $m^*$, we compute an
$(M_{\mathcal{D}_f},m[k^2;\delta'])-$admissible point $m^*$, with $\delta'=\delta\cdot\frac{\lceil k/2\rceil}{k^2}$, instead. Then, the same argument\footnote{The argument in~\cite{Sagraloff14} only uses that, in each iteration, we choose an arbitrary point $m^*\in [m-\lceil k/2\rceil \cdot \delta,m+\lceil k/2\rceil \cdot \delta]$.} as in~\cite{Sagraloff14} yields:

\begin{thm}\label{refine2}
For refining $I_0$ to a size less than $2^{-L}$, the algorithm \textsc{NewRefine$^*$} needs $O(k\cdot (\log n+\log(\tau+L)))$ iterations. In each 
iteration, we need to compute a constant number of
$(M_{\mathcal{D}_f},m[k^2;\delta'])-$admissible points $m^*$, with $m[k^2;\delta']\subset I_0$ and $\delta'=2^{-O(\tau+\log n+L)}$. In addition, the polynomials $f$ and $f'$ have to evaluated at $m^*$ to an absolute 
precision bounded by $O(\log \max(1,|f(m^*)|^{-1})+\log n+L+\tau)$.
\end{thm}

Combining Theorems~\ref{refine2} and~\ref{thm:Let--be-1}, we obtain a bound on the complexity of refining $I_0$ to a size less than $2^{-L}$:

\begin{cor}\label{cor:bitcomplex:refine}
For refining $I_0$ to a size less than $2^{-L}$, the algorithm \textsc{NewRefine$^*$} needs
\[
\tilde{O}(k^5\cdot(k+\log n)\cdot\log n\cdot(k\log n+\tau+L+n\log\max(1,b_0)))
\]
bit operations. For each endpoint $p$ of the interval returned by \textsc{NewRefine}, it holds that
\[
M_{\mathcal{D}_f}(p)=2^{-O(\ell+k(k\log n+\tau+L+n\log\max(1,b_0)))}. 
\]
with $\ell:=\log \min(1,M_{\mathcal{D}_f}(a_0),M_{\mathcal{D}_f}(b_0))^{-1}$. 
\end{cor}

\section{Computing a Weak Covering}

\label{sec:Computing-a-Weak}

We now describe how to compute a weak \emph{ $(L,[0,1+1/n])$-}covering
for a given $(n,k,\tau)$-nomial $f$ in polynomial time. We first compute an upper bound $\tilde{\tau}\in\mathbb{Z}$
for $\tau$ with $\tau\le\tilde{\tau}\le \tau+2$, and define $\delta:=\min(2^{-2\tau-2},1/n)\cdot k^{-2}$. Then, in the first step, we 
compute $(M_{\mathcal{D}_f},m[k^2;\delta])-$admissible points $a^*$ and $b^*$ for 
$m:=2^{-2\tau-2}$ and $m:=1+2/n$, respectively.
Then, we follow the approach as outlined in the first part of Section~\ref{sec:overview} to compute a weak $(L,[a^*,b^*])$-covering
for $f$, where we use the algorithm \textsc{NewRefine$^*$} from the previous Section to refine isolating intervals
for the roots of the fractional derivatives of $f$ to a size less than $2^{-L}$. The so obtained covering is indeed
also a weak \emph{ $(L,[0,1+1/n])$-}covering for $f$, which follows from the fact that $b^*\ge 1+1/n$ and each positive root of $f$ is lower bounded by 
$(1+\max_{i=1}^k |f_i|/|f_1|)^{-1}$ due to Cauchy's root bound~\cite{Yap:book}. For details, consider the exact definition
of Algorithm~\ref{alg:Main}.
\begin{algorithm}[H]
\SetKwInOut{Input}{Input}
\SetKwInOut{Output}{Output}
\SetKwInOut{Assumption}{Assumption}

\Input{An $(n,k,\tau)$-nomial $f$ and a non-negative integer $L\in\N$.}

\Output{A weak $(L,[0,1+1/n])$-covering of $f$.}\medskip

Compute $\tilde{\tau}\in\mathbb{N}$ with $\tau\le\tilde{\tau}\le\tau+2$.\medskip

$\delta:=\frac{1}{k^2}\cdot\min(\frac{1}{n},2^{-2\tilde{\tau}-2})$

Compute $(M_{\mathcal{D}_f},m[k^2;\delta])-$admissible points $a^*$ and $b^*$ for $m:=2^{-2\tilde{\tau}-2}$ and $m:=1+\frac{2}{n}$, respectively. Compute the sign of $f$ at $x=a^*$ and $x=b^*$.\medskip

\For{$i=k-1$ to $0$}{ 

\eIf{$i=k-1$}{

Compute a trivial weak $(L,[a^*,b^*])$-covering $W_{\ensuremath{k-1}}$
for $\f f{k-1}$ ($\f f{k-1}$ has only one monomial).

$W_{k-1}=\{(a^*,a^*),(b^*,b^*)\}$.

}{

$W_{i+1}$ = weak $(L,[a^*,b^*])$-covering for $\f f{i+1}$ computed
in the previous iteration of this loop.

$W_{i}=W_{i+1}$.

\For{each consecutive intervals $(a,b)$ and $(c,d)$ in $W_{i+1}$
}{

Compute signs of $\f fi(b)$ and $\f fi(c)$

\If{$f^{(i)}(b)f^{(i)}(c)<0$}{

Use \textsc{NewRefine$^*$} to refine the isolating interval $(b,c)$ to a new interval $(b^{\prime},c^{\prime})$
of length at most $2^{\ensuremath{-L}}$.

Compute signs of $\f fi(b^\prime)$ and $\f fi(c^\prime)$.

$W_{i}=W_{i}\cup(b^{\prime},c^{\prime})$

}

}

}

}

\Return $W_{0}$.

\caption{\label{alg:Main}Compute a weak $(L,[0,1])$-covering of $f$}
\end{algorithm}

Correctness of the algorithm follows directly from our considerations in Section~\ref{sec:overview}. 
Further notice that, for each $i$ in the outermost for-loop of the algorithm, we add at most $k-i-1$ intervals to $W_i$
to obtain $W_{i+1}$ as $f^{[i]}$ has at most $k-i-1$ positive real roots. Hence, each list $W_i$ contains at most $k^2$
many intervals. It remains to bound the running time of Algorithm~\ref{alg:Main}. The proof of the following Lemma follows
in a straight forward manner from Theorem~\ref{thm:Let--be-1}, Corollary~\ref{cor:bitcomplex:refine}, and the fact that we need to call the refinement algorithm at most $k$ times for each fractional derivative.

\begin{restatable}{lem}{weakcover}
\label{complex:weakcover}
Algorithm \ref{alg:Main} computes a weak $(L,[0,1+\frac{1}{n}])$-covering for $f$ consisting of at most $k^2$ many intervals. Its bit complexity is
$\tilde{O}(k^7(k+\log n\cdot(k\log n+\tau+L)\cdot )\log n).$
\end{restatable}

\begin{proof*}
According to Theorem~\ref{thm:Let--be-1}, the cost for computing $a^*$ and $b^*$ is bounded by $\tilde{O}(k^4(k+\log n)(k\log n+\tau))$. 
The refinement algorithm is called at most $k^2$ many times for refining the roots of the fractional derivatives. Corollary~\ref{cor:bitcomplex:refine} thus yields a bound of size
$\tilde{O}(k^7\cdot(k+\log n)(k\log n+\tau))$ for the bit complexity of the refinements. The computations of the signs of the factional derivative $f^{[i]}$ at the endpoints of the intervals in $W_{i+1}$ is dominated by this bound as the refinement algorithm returns intervals whose endpoints are admissible with respect to $M_{\mathcal{D}_f}$. Thus, the computation of each such admissible point already yields the sign of all fractional derivative at this point. 
\end{proof*}

In order to further process a weak $(L,[0,1+1/n])$-covering for $f$, we need the intervals in the 
weak covering to be well separated. For given $L,\lambda\in\N_0$, we say that a list $\mathcal{L}$ of intervals 
is \emph{$(L,\lambda)$-separated} if the distance $\operatorname{dist}(I,J)$ between $I$ and its neighboring
intervals is at least $\min(2^{-L},\lambda\cdot w(I))$. Notice that, starting from an arbitrary list $\mathcal{L}$ of intervals,
we can always deduce an $(L,\lambda)$-separated list $\mathcal{L}'$ from $\mathcal{L}$ in a way such that each interval in $\mathcal{L}$ is contained in an interval from $\mathcal{L}'$. Namely,
this can be achieved by recursively merging pairs of intervals $I,J\in\mathcal{L}$ that violate the above condition until 
the actual list is \emph{$(L,\lambda)$-separated}. It is easy to see that
\[
w(\mathcal{L}')\le (2+\lambda)^{|\mathcal{L}|}\cdot \max(2^{-L},w(\mathcal{L})),
\]
where $w(\mathcal{L})$ and $w(\mathcal{L}')$ denote the maximal width of an interval in $\mathcal{L}$ and $\mathcal{L}'$, respectively.
Hence, by first computing a weak $(L',[0,1+1/n])$-covering $\mathcal{L}$, with $L'=L+k^2\cdot\log(2+\lambda)$ and $|\mathcal{L}|=O(k^2)$, and then recursively merging the intervals, we obtain a weak $(L,[0,1+1/n])-$covering for $f$ that is also 
$(L,\lambda)$-separating and whose intervals have width at most $2^{-L}$.
From Lemma~\ref{complex:weakcover}, we thus conclude:
\begin{cor}
For any $\lambda,L\in\N_0$, we can compute a $(L,\lambda)$- separating weak $(L,[0,1+1/n])-$covering for $f$ in $
\tilde{O}(k^7(k+\log n)\cdot(k\log n+\tau+L+k^2\log(2+\lambda))\cdot \log n)
$
bit operations.
\end{cor}

\section{$T_{l}$-test}

In the previous section, we have shown how to compute a weak $(L,[0,1+1/n))$-covering of
a given $(n,k,\tau)$-nomial $f$. Now, we aim to convert this weak
covering to a covering of $f$. For this, we need an algorithm to
count the number of roots of $f(x)$ contained in a given disk.
Recent work~\cite{BeckerS0Y15} introduces a simple corresponding algorithm, denoted $T_{l}$-test, which is based on Pellet's Theorem. More precisely, for an arbitrary polynomial $F\in\C[x]$, a disk $\Delta=\Delta_r(m)\subset\C$, and a parameter $K\ge 1$, we consider the 
inequality
\begin{align}\label{Tltest}
T_{l}(\Delta,K,F):\left\vert \frac{\d Fl(m)r^{l}}{l!}\right\vert -K\cdot\sum_{i\neq l}\left\vert \frac{\d Fi(m)r^{i}}{i!}\right\vert >0.
\end{align}
Hence, we check whether the absolute value of the $l$-th coefficient $a_l$ of
$F_{\Delta}(x)=f(m+rx)=\sum_{i=0}^n a_ix^i$ dominates the sum of the absolute values of
all remaining coefficients weighted by the parameter $K$. We say that
$T_{l}(\Delta,K,F)$ succeeds if the above inequality is fulfilled. Otherwise, we say that it fails. In case of success (for any $K\ge 1$), $\Delta$ contains exactly $l$ roots of $F$ counted with multiplicity, whereas we have no information in case of a failure. However, in~\cite{BeckerS0Y15}, we derive sufficient conditions on the success of the $T_l$-test:
\begin{thm}\label{successTltest}
[\cite{BeckerS0Y15}, Corollary 1] Let $F\in\C[x]$ be a polynomial of degree
$n$, and $\Delta_{r}(m)$ be a disk. If $\Delta_r(m)$ as well as the enlarged
disk $\Delta_{256n^{5}r}(m)$ contain $l$ roots of $F$ counted with multiplicity, then
\textup{$T_{l}(\Delta_{16nr}(m),\frac{3}{2}, F)$ succeeds.}
\end{thm}
Unfortunately, the above test has two major drawbacks when dealing with sparse polynomials. First, we need to compute the coefficients $F_{\Delta}$ exactly, which we cannot afford as the bitsize of each coefficient is at least linear in $n$. Second, an even more severe, there are $n$ coefficients to be computed. Hence, using the above approach directly to count the number of roots of a sparse polynomial $f$ does not work. Instead, we propose two modifications to overcome these issues. The first modification, namely to use approximate (in a proper manner) instead of exact arithmetic, has already been considered in previous work. However, the second modification is more subtle. It exploits the fact that, for a suitably chosen disk centered at some admissible point, only the first $k^2$ coefficient are relevant for the outcome of the above test.

We first go into details with respect to our first modification.
Let us define $E_{\ell}:=|a_l|$
and $E_r:=K\cdot\sum_{i\neq l}|a_i|$ the expressions on the left and right hand side of the inequality in (\ref{Tltest}). We aim to check whether $E_{\ell}-E_{r}>0$ or not. In general, if a predicate $\mathcal{P}$ is of the latter form $\mathcal{P}=(E_{\ell}-E_r>0)$ with two (computable) expressions $E_{\ell}$ and $E_r$, you can compute approximations $\tilde{E}_{\ell}$ and $\tilde{E}_r$ of $E_{\ell}$ and $E_{r}$ with $|\tilde{E}_{\ell}-E_{\ell}|<2^{-L}$ and $|\tilde{E}_{r}-E_{r}|<2^{-L}$ for $L=1,2,4,\ldots$ For a certain $L$, you may then try to 
compare $E_{\ell}$ and $E_r$ taking into account their corresponding approximations and the approximation error. Eventually (i.e. for a sufficiently large $L$), you either succeed, in which case you can return the sign, or assert that $E_{\ell}$ and $E_r$ are good approximations of each other. In the latter case, you just return a flag called Undecided. In short, this is the idea of so-called \emph{soft-predicates}. 
For details, we refer to~\cite{BeckerS0Y15}. 

\begin{algorithm}[H]
\SetKwInOut{Input}{Input}
\SetKwInOut{Output}{Output}\Input{ A predicate $\mathcal{P}$ defined by non-negative expressions
$E_{\ell}$ and $E_{r}$ , with $E_{\ell}\neq0$ or $E_{r}\neq0$;
i.e. $\mathcal{P}$ succeeds if and only if $E_{\ell}>E_{r}$. A rational constant
$\delta>0$. }

\Output{True, False, or Undecided. In case of True (False), $\mathcal{P}$
succeeds (fails). In case of Undecided, we have $\frac{1}{1+\delta}\cdot E_{\ell}<E_{r}\leq(1+\delta)\cdot E_{\ell}$}
\caption{\label{alg:-soft}Soft Predicate $\tilde{\mathcal{P}}$}
\end{algorithm}

Notice that, in cases where $E_{\ell}$ considerably differs from $E_r$, the soft predicate $\tilde{\mathcal{P}}$ allows us to compute the
sign of $\mathcal{P}$ without the need of exact
arithmetic. In all other cases (i.e. if it returns Undecided), we know at least that $E_{\ell}$
and $E_{r}$ are good approximations of each other. 
We remark that, in~\cite{BeckerS0Y15}, the above soft predicate $\tilde{\mathcal{P}}$ was only
described for $\delta=\frac{1}{2}$, however, it easily generalizes to any constant $\delta$. In~\cite[Lem.~2]{BeckerS0Y15}, it has been shown that, for any constant $\delta$, Algorithm~\ref{alg:-soft} needs an $L_{0}$-bit approximation of \textup{$E_{\ell}$
}and $E_{r}$ with $L_{0}$ bounded by
\[
L_{0}\leq2\cdot(\max(1,\log\max(E_{\ell},E_{r})^{-1})+4).
\]

In~\cite{BeckerS0Y15}, we considered a soft-variant of the $T_l$-test, where we compared the expressions $E_{\ell}:=|a_{l}|$
and $E_r:=\sum_{i\neq l}|a_i|$. Now, we apply the above soft-predicate to the expressions $E_{\ell}:=a_l$
and $E_r:=\sum_{i\neq l}^{i\le k^2}|a_i|$, that is, we replace the entire sum $\sum_{i\neq l}|a_i|$ by its truncation after the first $k^2$ terms. However, we will make the assumption that the truncated sum is upper bounded by $\frac{|a_0|}{128}$; see Algorithm~\ref{alg:-test}. This might look haphazardly at first sight, however, we will later see that the latter condition is always fulfilled for a $k$-nomial $F$ and a suitable disk $\Delta_r(m)$ centered at an admissible point. 
\begin{algorithm}[H]
\SetKwInOut{Input}{Input}
\SetKwInOut{Output}{Output}
\SetKwInOut{Assumption}{Assumption}\Input{ An $(n,k,\tau)$-nomial $f(x)$, a disk $\Delta:=\Delta_{r}(m)$
in the complex space and an integer $l$ with $0\le l\le k$. It is required that $\sum_{i>k^{2}}|a_{i}|\leq\frac{|a_{0}|}{128}$, where
$f_{\Delta}(x)=\sum_{i=0}^{n}a_{i}\cdot x^{i}$.}

\Output{True or False. If the algorithm returns True then the disk
$\Delta_{r}(m)$ contains exactly $l$ roots. }
Define $E_{\ell}\eqdef|a_{l}|$ and $E_{r}\eqdef\frac{65}{64}\cdot\sum_{i\neq l}^{i\leq k^{2}}|a_{i}|$. 

Define predicate $\mathcal{P}=(E_{\ell}-E_{r}>0)$.

\Return output of Algorithm \ref{alg:-soft} on predicate $\mathcal{P}$
with $\delta=\frac{1}{128}$.

\caption{\label{alg:-test}$\tilde{T}_{l}$-test}
\end{algorithm}

\begin{restatable}{lem}{Tltest}
\label{lem:For-a-disk}For a disk $\Delta:=\Delta_{r}(m)\subset\mathbb{C}$, the $\tilde{T}_{l}$-test needs to compute $L$-bit approximations of $E_{\ell}$ and $E_r$ with   
$
L\le L(m,r,f):=2\cdot\left(5+\log n-\log\max_i |a_i|\right).
$
If $T_{l}(\Delta,\frac{3}{2},f)$
succeeds, then the $\tilde{T}_{l}$-test returns True. Running Algorithm \ref{alg:-test} for all $l=0,\ldots,k$ uses a number of bit operations upper bounded by
$
\tilde{O}(k^{2}\cdot(k+\log n)(L(m,r,f)+\tau+n\log\max(1,m)+
k^2\cdot(\log n +\log\max(1,r)))).$
\end{restatable}

\begin{proof*}
From the assumption, it follows that $\max_{i=0,\ldots,n} |a_i|=\max_{i=0,\ldots,k^2} |a_i|\le\frac{1}{2}\cdot\max(|E_{\ell}|,|E_r|)$. This yields the claimed bound
on the absolute error to which $E_{\ell}$ and $E_r$ need to be computed. 
We now prove correctness. If the algorithm returns True, then $E_{\ell}>E_{r}$, and thus
$|a_{l}|>\frac{65}{64}\cdot\sum_{i\neq l}^{i\leq k^{2}}|a_{i}|.$ If $l=0$, then 
$\sum_{i\neq 0}|a_{i}|<\frac{64}{65}\cdot |a_0|+\frac{1}{128}\cdot |a_0|<|a_0|$. Otherwise, 
we have $|a_{l}|>\frac{65}{64}\cdot\sum_{i\neq l}^{i\leq k^{2}}|a_{i}|\ge
\sum_{i\neq l}^{i\leq k^{2}}|a_{i}|+\frac{1}{64}\cdot |a_0|\ge \sum_{i\neq
l}^{i\leq n}|a_{i}|$. Hence, in both cases, $T_l(\Delta,1,f)$ succeeds, which implies that $\Delta$ contains exactly $l$ roots.

Now, suppose that $T_{l}(\Delta,\frac{3}{2},f)$ succeeds. If the $\tilde{T}_{l}$-test returns
Undecided, then $\frac{128}{129}\cdot E_{\ell}<E_{r}\leq\frac{129}{128}\cdot E_{\ell}$. 
On the other hand, we have $|a_l|>\frac{3}{2}\sum_{i\neq l}^{\le n}|a_i|\ge \frac{3}{2}\sum_{i\neq l}^{\le k^2}|a_i|$, and thus
$E_{\ensuremath{\ell}}>\frac{3}{2}E_{r}$, which contradicts the fact that $\frac{128}{129}\cdot E_{\ell}<E_{r}$.
If the $\tilde{T}_{l}$-test returns False,
a similar argument yields a contradiction as well. This shows that success of $T_l$ implies that $\tilde{T}_l$ returns True.
It remains to show the claimed bounds on the bit complexity. It suffices to estimate the cost for computing 
an $L(m,r,f)$-bit approximations of $E_{\ell}$ and $E_{r}$. The $i$-th coefficient
$a_{i}$, with $i\le k^2$, can be computed
by evaluating the $(n,k,\tau+k^2\cdot(\log n+\log\max(1,r)))$-nomial $g_i=f^{(i)}(x)r^{i}/i!$
at $x=m$. In order to compute $L(m,r,f)$-bit approximations of $E_{\ell}$ and $E_{r}$, we need to compute
an $(L(m,r,f)+2\log k)$-bit approximation of each $g_i(m)$, for $i=0,\ldots,k$. According to Lemma
\ref{lem:Let--be}, this can be done using $\tilde{O}(k^{2}\cdot(k+\log
n)(L(m,r,f)+n\log\max(1,m)+\tau+k^2\cdot(\log n+\log\max(1,r)))$ bit operations.
\end{proof*}
\medskip
Notice that, in order to actually use
the $\tilde{T}_l$-test for counting the roots in a disk $\Delta$, we need two conditions to be satisfied.
First, we need the condition $\sum_{i>k^{2}}|a_{i}|\leq\frac{|a_{0}|}{128}$
to be true. Second, we need to satisfy the preconditions of the $T_{l}$-test.

\begin{thm}\label{thm:Let--be-2}Let $f$ be a $(n,k,\tau)$-nomial as in (\ref{eq:defsparse}), let
$\Delta:=\Delta_{r}(m)$ be a disk centered at some $m\in\R_{>0}$ with
$\frac{m}{r}>n^{16}$, and let \textup{$f_{\Delta}(x)=\sum_{i=0}^{n}a_{i}\cdot x^{i}$}. Further suppose that $\Delta_{\frac{r}{k^{4k+2}}}(m)$ does not 
contain any roots of $f$. Then, it holds that $\sum_{i>k^{2}}|a_{i}|\leq\frac{|a_{0}|}{128}$.
\end{thm}

\begin{proof*}
Let $z_{1},z_{2},\ldots,z_{n}$ be the complex roots of $F(x)$, then
$
\frac{a_{i}}{a_{0}}=\frac{F^{(i)}(m)}{F(m)\cdot i!}\cdot r^{i}=\frac{r^{i}}{i!}\cdot\sum_{(j_{1},j_{2},\ldots,j_{i})}\frac{1}{\prod_{\ell=1}^{i}(m-z_{j_{\ell}})},
$
where we sum over all tuples $(j_{1},j_{2},\ldots,j_{i})$ with distinct entries $j_s$, $1\le j_s\le n$.
For a fixed tuple $(j_{1},j_{2},\ldots,j_{i})$, at most $k$ of the $i$ roots $z_{\ensuremath{j_{1}}},z_{\ensuremath{j_{2}}},\ldots,z_{\ensuremath{j_{i}}}$ can appear in the corresponding term of the above sum. At
most $k$ of these roots are contained in the code $C_{n}$ as defined in Figure \ref{fig:Cone-1}, whereas the remaining
$i-k$ roots are located outside of $C_{n}$. Since $\frac{m}{r}>n^{16}$, the distance from $m$ to any of these roots is at least $n^{15}r$.
Also, since $\Delta_{\frac{r}{k^{4k}}}(m)$ does not contain any
roots of $F(x)$, distance of $m$ from the roots in $C_{n}$ is at least
$\frac{r}{k^{4k}}$. Thus, we get 
$
\sum_{(j_{1},j_{2},\ldots,j_{i})}\frac{1}{\prod_{\ell=1}^{i}|m-z_{j_{k}}|}\leq{n \choose i}\cdot\frac{k^{4k^2}}{r^{k}\cdot(n^{5}r)^{i-k}}.
$
Hence, for $i>k^2$, we get
\begin{align*}
\frac{|a_i|}{|a_0|} & \leq\frac{r^{i}}{i!}\cdot{n \choose i}\cdot\frac{k^{4k^2+2k}}{r^{k}\cdot(n^{15}r)^{i-k}}=\frac{1}{i!}\cdot{n \choose i}\cdot\frac{k^{4k^2+2k}}{n^{15(i-k)}}\\
 & \leq\frac{1}{i!\cdot i!}\cdot\frac{k^{4k^2+2k}}{n^{14i-15k}} \leq\frac{1}{i!\cdot i!}\cdot\frac{k^{4k^{2}+2k}}{n^{6i}}\tag{By using the fact that \ensuremath{{n \choose i}\le\frac{n^{i}}{i!}} and $15k<8k^2<8i$}\\
 & \leq\frac{1}{5!\cdot n\cdot i!}\cdot\frac{k^{4k^{2}+2k}}{k^{6k^2}} \leq\frac{1}{120\cdot n\cdot i!}\cdot\left(\frac{1}{k^{2}}\right)^{k^2-2k}<\frac{1}{128n}
\end{align*}
Hence, summing up over all $i>k^2$ proves the claim.
\end{proof*}

\medskip

The following Corollary is now an immediate consequence of the above theorem and Lemma \ref{thm:Let--be-2}.

\begin{cor}
\label{cor:Let--be}Let $f(x)\in\R[x]$ be as in $(n,k,\tau)$-nomial as in (\ref{eq:defsparse}).
Let $m,r\in\R^{+}$. Let $m^{*}$ be a $(M_{D_{f}},m[k^{2};\frac{r}{k^{2}}])$
-admissible point and $r^{*}=2r$. Define $\Delta=\Delta_{r^{*}}(m^{*})\supseteq\Delta_{r}(m)$ and $f_{\Delta}(x)=\sum_{i=0}^{n}a_{i}\cdot x^{i}$. Further assume that $\frac{m}{r}\geq2(1+n^{16})$, then $\sum_{i>k^{2}}|a_{i}|\leq\frac{|a_{0}|}{128}.$
\end{cor}

In the next step, we show how to satisfy the precondition of the $T_{l}$-test.
Theorem~\ref{successTltest} says that if $\Delta_{256n^{5}r}(m)$
does not contain any of the roots which are not contained in $\Delta_{r}(m)$, then
$T_l(\Delta_{16nr},f)$ succeeds for some $l$.
Let us define $M=256n^{5}r$, and let $\Delta_i:=\Delta_{M^{i}r}(m)$ for $i=0,1,\ldots,k+1$.
Further assume that $r$ has been chosen sufficiently small enough such that each of disks is contained in the cone $C_n$. Since $C_n$ contains at most $k$
roots, there must exist a $j$ with $0\leq j\leq k$ such that $\Delta_{j+1}-\Delta_{j}$
does not contain any root. Hence the $T_{l}$-test will succeed on $\Delta_{16nM^{j}r}(m)$.
So instead of running the $T_{l}$-test on some initial disk $\Delta_{r}(m)$, we run
it on all disks $\Delta_{16nM^{i}r}(m)$ for $i=0,1,\ldots,k$, and return the first
disk on which the $T_{l}$-test succeeds; see Algorithm \ref{alg:-test-1}.

Correctness of the algorithm follows immediately from the above considerations. The condition on $m$ and $r$ guarantees that each of the disks $\Delta_i$ is contained in $C_n$. Lemma~\ref{lem:Algorithm--returns} gives a bound on its running time.

\begin{algorithm}
\SetKwInOut{Input}{Input}
\SetKwInOut{Output}{Output}
\SetKwInOut{Assumption}{Assumption}\Input{ A $(n,k,\tau)$-nomial $f(x)$, a disk $\Delta:=\Delta_{r}(m)$
in the complex space. We assume $m\geq r+2Rnr$ with $R=2^{8k+4}n^{5k+16}$.}

\Output{ A disk $\Delta_{r'}(m')$ such that $\Delta_{r}(m)\subseteq\Delta_{r^{\prime}}(m^{\prime})$
along with number of roots of $f(x)$ contained in $\Delta_{r^{\prime}}(m^{\prime})$
}
\begin{enumerate}
\item Compute an $(M_{D_{f}},m[k^{2};\frac{r}{k^{2}}])$-admissible point
$m^{*}$.
\item Let $m^{\prime}=m^{*}$ and $r^{\prime}=2r$.
\item Let $M=256n^{5}r^{\prime}$. 
\end{enumerate}
\For{each $0\leq i\leq k$}{ 
\For{each $0\leq l\leq k$}{ 
Perform the $\tilde{T_{l}}$-test, that is Algorithm \ref{alg:-test},
on $\Delta_{16nM^{i}r^{\prime}}(m^{\prime})$.
\If{$\tilde{T_{l}}$-test succeeded in the previous step}{
\Return $\Delta_{16nM^{i}r^{\prime}}(m^{\prime})$ and $l$.
}
}
}
\caption{\label{alg:-test-1}Wrapper $\tilde{T}_{l}$-test}
\end{algorithm}

\begin{restatable}{lem}{growingdisks}
\label{lem:Algorithm--returns}Algorithm \ref{alg:-test-1} returns
a disk $\Delta_{r^{\prime}}(m^{\prime})$, with $r^{\prime}\le Rr$ and $m-r\leq m^{\prime}\leq m+r$, together with the number of roots
of $f(x)$ in $\Delta_{r'}(m')$. Its bit complexity is bounded by \textup{$\tO(k^{5}\cdot(k+\log n)\cdot(k^{2}\log n+n\log\max(1,|m|)+\tau+\log\frac{1}{r}))$}.
\end{restatable}

\begin{proof*}

The condition $m\geq r+2Rnr$ with $R=2^{8k+4}n^{5k+16}$ implies that
all the disks considered in the Algorithm \ref{alg:-test-1} are contained in
the cone $C_{n}$. In addition, the condition of Corollary
\ref{cor:Let--be} is fulfilled. 

One iteration of the inner for loop uses a number of bit operations bounded by $\tilde{O}(k^{2}\cdot(k+\log n)\cdot (L(m^{\prime},R^{\prime},f)+\tau+n\log\max(1,m^{\prime})+k^{2}(\log n+\log\max(1,R^{\prime}))$; see Lemma~\ref{lem:For-a-disk}. Here, $R^{\prime}\leq R$ and
$m-r\leq m^{\prime}\leq m+r$. In addition, $L(m^{\prime},R^{\prime}):=L(m^{\prime},R^{\prime},f):=2\cdot\left(5+\log n-\log\|f_{\Delta}\|_{\infty}\right).$

If $f_{\Delta}(x)=\sum_{i=0}^{n}a_{i}\cdot x^{i}$, then obviously
$\|f_{\Delta}\|_{\infty}\geq |a_{0}|=|f(m^{\prime})|$.
Since $m^{\ensuremath{\prime}}$ is an $(M_{D_{f}},m[k^{2};\frac{r}{k^{2}}])$-admissible
point, Lemma \ref{lem:Let--be-2} implies that \\ $|M_{D_{f}}(m^{\prime})|\geq2^{-O(k(\tau+n\log\max(1,m^{\prime})+k\log n+\log\max(1,\frac{k^{2}}{r})))}$.
Thus, we conclude that $-\log\|f_{\Delta}\|_{\infty}\leq k(\tau+n\log\max(1,m^{\prime})+k\log n+\log\frac{1}{r})$, and $L(m^{\prime},R^{\prime})\leq O(k(\tau+n\log\max(1,m^{\prime})+k\log n+\log\frac{1}{r}))$.
It follows that Algorithm \ref{alg:-test-1} runs in time $\tilde{O}(k^{4}\cdot(k+\log n)(k(\tau+n\log\max(1,m^{\prime})+k\log n+\log\frac{1}{r})+\tau+k^{2}(\log\max(1,R)+\log n)+n\log\max(1,m^{\prime})))=\tO(k^{5}\cdot(k+\log n)\cdot(k^{2}\log n+n\log\max(1,m^{\prime})+\tau+\log\frac{1}{r}))$.
\end{proof*}

\section{Computing a Covering}\label{sec:covering}

We now show to compute an $(L,[0,1+1/n])$-covering from a weak $(L',[0,1+\frac{1}{n}])$-covering,
For this, we apply Algorithm \ref{alg:-test-1} to the one-circle regions of the intervals in the weak covering.
The following Lemma shows that the requirements in Algorithm \ref{alg:-test-1} are fulfilled if we choose $L'$ large enough.
In addition, by ensuring that the intervals in the weak covering are well separated from each other, we can ensure that
the corresponding disks returned by Algorithm \ref{alg:-test-1} are disjoint.

\begin{algorithm}
\SetKwInOut{Input}{Input}
\SetKwInOut{Output}{Output}
\Input{ An $(n,k,\tau)$-nomial $f(x)$ and a positive integer $L$.} 
\Output{ An $(L,[0,1+1/n])$-covering for $f$.
}
\begin{enumerate}
\item Let $R:=2^{8k+4}n^{5k+16}$ and $L^{\prime}=L+\lceil\log R\rceil+4\tau+5$.
Compute a weak $(L^{\prime},[0,1+\frac{1}{n}])$-covering $\mathcal{L}$
for $f$ that is $(L^{\prime},8R)$-separated.\vspace{-0.1cm}
\item $\mathcal{L}^{\prime}=\emptyset$\vspace{-0.1cm}
\end{enumerate}
\For{each interval $I=(a,b)\in\mathcal{L}$}{ \vspace{-0.1cm}
\begin{enumerate}
\item $\Delta=\Delta_{\frac{b-a}{2}}(\frac{a+b}{2})$=one circle region of $I$.\vspace{-0.1cm}
\item $(\Delta_{r^{\prime}}(m^{\prime}),\mu)$= output of Algo.\vspace{-0.1cm}
\ref{alg:-test-1} on $f$ and $\Delta$.
\item $\mathcal{L}^{\prime}=\mathcal{L}^{\prime}\cup\{(\Delta_{r^{\prime}}(m^{\prime}),\mu)\}$\vspace{-0.1cm}
\end{enumerate}
}
\Return$\mathcal{L}^{\prime}$. 
\caption{\label{alg:-strongcover}Computing a $(L,[0,1+\frac{1}{n}])$-covering}
\end{algorithm}

\begin{restatable}{lem}{Final}
\label{lem:Algorithm--0,1}Algorithm \ref{alg:-strongcover} computes
an $(L,[0,1+\frac{1}{n}])$-covering $\mathcal{L^{\prime}}$ for $f$
using $\tO(k^{7}\cdot(k+\log n)(k^3\log n+\tau+L))$ bit operations. The distance
between any two disks of $\mathcal{L^{\prime}}$ is at least $32\cdot2^{-L}$, and
$\Delta\cap\R\subset (2^{-3\tau},2)$ for any disk $\Delta$ in $\mathcal{L}^{\prime}$.
\end{restatable}

\begin{proof*}
The output $\mathcal{L}^{\prime}$ surely covers all the real roots
of $f$ in the interval $[0,1+\frac{1}{n}]$. Since the weak covering $\mathcal{L}$
computed in Algorithm \ref{alg:-strongcover} is $(L^{\prime},8R)$-separated
and since Algorithm \ref{alg:-test-1} only blows up any disk by a factor
of $R$, we conclude that disks in $\mathcal{L^{\prime}}$ are still separated
by at least $4R2^{-L^{\prime}}\geq32\cdot2^{-L}$. In addition, the radius
of each disks in $\mathcal{L^{\prime}}$ is at most $R2^{-L^{\prime}}\geq2^{-L}$. 

Notice that the left endpoint of any interval in $\mathcal{L}$ is at least
$2^{-2\tau-3}$. Thus, for any disk $\Delta$ from $\mathcal{L}^{\prime}$ the left endpoint of the interval $\Delta\cap\R$
is at least $2^{-2\tau-3}-R2^{-L^{\prime}}\geq2^{-2\tau-5}$. A similar
argument yields the claimed bound on the right end points of $\Delta\cap\R$.

The running time bounds follow from the stated upper bound on $L^{\prime}$
and $R$ and the fact that $m^{\prime} \leq 1 + O(\frac{1}{n}+2^{-L}))$ is
always satisfied.
\end{proof*}

It remains to show how to compute an $(L,[0,\infty))$-covering for $f$ from an $(L,[0,1+\frac{1}{n}))$-covering $\mathcal{L}_1$ for $f$ and 
an $(L,[0,1+\frac{1}{n}))$-covering $\mathcal{L}_2$ for $x^nf(\frac{1}{x})$. We first derive an $(L,[\frac{n}{n+1},\infty))$-covering for 
$f$ from $\mathcal{L}_2$ by inverting the disks $\Delta$ in $\mathcal{L}_2$. The proof of the following lemma is straight forward.
\begin{lem}
\label{lem:Let--be-3}Let $\mathcal{L}$ be an $(L,[0,1+\frac{1}{n}])$-covering
of $x^{n}f(\frac{1}{x})$ as computed by Algorithm \ref{alg:-strongcover}, and
$\mathcal{L}^{\prime}:=\{(\Delta^{-1},\mu):(\Delta,\mu)\in\mathcal{L}\}$ be the list obtained from 
$\mathcal{L}$ by inverting the disks in $\mathcal{L}$ (i.e. $\Delta_r(m)^{-1}=\Delta_{r'}(m')$ with $r'=\frac{2r}{m^2-r^2}$ and 
$m'=\frac{m}{m^2-r^2}$). Then, $\mathcal{L}^{\prime}$ is an  $(L^{\prime},[\frac{n}{n+1},\infty))$-covering
of $f$ with $L^{\prime}\geq L-6\tau$ and the distance between two disks
in $L^{\prime}$ is at least $8\cdot 2^{-L}$.
\end{lem}

Finally, we merge an $(L,[0,1+1/n))$-covering $\mathcal{L}_1$ and 
an $(L,[\frac{n}{n+1},\infty))$-covering $\mathcal{L}_2$ for $f$. Here, we assume that $L>3+\log n$, and that the coverings are computed
using Algorithm~\ref{alg:-strongcover} and by inverting the $(L,(0,1+1/n))$-covering for $x^n\cdot f(1/x)$ to obtain $\mathcal{L}_2$. 
This guarantees that the distance between any two disks in either $\mathcal{L}_1$ or $\mathcal{L}_2$ is at least $8\cdot 2^{-L}$.
For the merging, we keep each disk from 
$\mathcal{L}_1$ that has no intersection with a disk from $\mathcal{L}_1$, and vice versa.
For each pair of elements $(\Delta_1,\mu_1)\in\mathcal{L}_1$ and $(\Delta_2,\mu_2)\in\mathcal{L}_2$ 
with $\Delta_1\cap\Delta_2\neq\emptyset$, we keep $(\Delta_1,\mu_1)$ (and omit $(\Delta_2,\mu_2)$) if the center of $\Delta_1$ is not larger than $1$. Otherwise, we keep $(\Delta_2,\mu_2)$ (and omit $(\Delta_1,\mu_1)$). Following this approach, 
we might loose some of the complex roots that are contained in the union of $\Delta_1$ and $\Delta_2$, 
however, we will not loose any real root. Thus, the so obtained list constitutes an $(L,(0,\infty))$-covering for $f$.

Notice that any two $(L,(0,\infty))$- and $(L,(-\infty,0))$-coverings for $f$ can be trivially merged by taking their union.
In addition, since the final covering contains a list of disjoint disks contained in the union of the cone $C_n$ and its reflection on the imaginary axis, and since the union of these two cones contains at most $2k-1$ roots of $f$, the number of disks is also bounded by $2k-1$. Hence, our main Theorem \ref{thm:main} follows. 
{
\bibliographystyle{plain}
\bibliography{ref}

\begin{thebibliography}{10}

\bibitem{DBLP:conf/issac/GarciaG12}
Maria~Emilia Alonso~Gar\c{c}ia and Andr{\'e} Galligo.
\newblock A root isolation algorithm for sparse univariate polynomials.
\newblock In {\em ISSAC}, pages 35--42, 2012.

\bibitem{Bastani11}
Osbert Bastani, Christopher~J. Hillar, Dimitar Popov, and J.~Maurice Rojas.
\newblock Randomization, {S}ums of {S}quares, {N}ear-{C}ircuits, and {F}aster
  {R}eal {R}oot {C}ounting.
\newblock {\em Contemp. Mathematics}, 556:145--166, 2011.

\bibitem{BeckerS0Y15}
Ruben Becker, Michael Sagraloff, Vikram Sharma, and Chee{-}Keng Yap.
\newblock A near-optimal subdivision algorithm for complex root isolation based
  on the pellet test and newton iteration.
\newblock {\em J. Symb. Comput.}, 2015.
\newblock In press.

\bibitem{Collins:1976}
George~E. Collins and R\"{u}diger Loos.
\newblock Polynomial real root isolation by differentiation.
\newblock In {\em SYMSAC}, pages 15--25, 1976.

\bibitem{Coste2005479}
Michel Coste, Tom\'{a}s Lajous-Loaeza, Henri Lombardi, and Marie-Francoise Roy.
\newblock Generalized {B}udan-{F}ourier theorem and virtual roots.
\newblock {\em J. Complexity}, 21(4):479 -- 486, 2005.

\bibitem{CUCKER1999}
F.~Cucker, P.~Koiran, and S.~Smale.
\newblock A polynomial time algorithm for diophantine equations in one
  variable.
\newblock {\em J. Symb. Comput.}, 27(1):21 -- 29, 1999.

\bibitem{Kerber2015377}
Michael Kerber and Michael Sagraloff.
\newblock Root refinement for real polynomials using quadratic interval
  refinement.
\newblock {\em Journal of Computational and Applied Mathematics}, 280:377 --
  395, 2015.

\bibitem{lenstra99}
Hendrik~W. Lenstra~(Jr.).
\newblock Finding small degree factors of lacunary polynomials.
\newblock {\em Number Theory in Progress}, 1:267--276, 1999.

\bibitem{McNamee-Pan}
J.M. McNamee and Victor~Y. Pan.
\newblock {\em Numerical Methods for Roots of Polynomials}.
\newblock Number~2 in Studies in Computational Mathematics. Elsevier Science,
  2013.

\bibitem{MSW-rootfinding2013}
K.~Mehlhorn, M~Sagraloff, and P.~Wang.
\newblock {From Approximate Factorization to Root Isolation with Application to
  Cylindrical Algebraic Decomposition}.
\newblock {\em J. Symb. Comput.}, 66(1):34 -- 69, 2015.

\bibitem{Pan:alg}
V.~Pan.
\newblock {Univariate Polynomials: Nearly Optimal Algorithms for Numerical
  Factorization and Root Finding}.
\newblock {\em J. Symb. Comput.}, 33(5):701--733, 2002.

\bibitem{Pan:2007}
Victor~Y. Pan, Brian Murphy, Rhys~Eric Rosholt, Guoliang Qian, and Yuqing Tang.
\newblock Real root-finding.
\newblock In {\em SNC}, pages 161--169, 2007.

\bibitem{Pan:2013}
Victor~Y. Pan and Elias~P. Tsigaridas.
\newblock On the boolean complexity of real root refinement.
\newblock In {\em ISSAC}, pages 299--306, 2013.

\bibitem{Rojas05}
J.~Maurice Rojas and Yinyu Ye.
\newblock On solving univariate sparse polynomials in logarithmic time.
\newblock {\em J. Complexity}, 21(1):87--110, 2005.

\bibitem{Sagraloff14}
Michael Sagraloff.
\newblock A near-optimal algorithm for computing real roots of sparse
  polynomials.
\newblock In {\em ISSAC}, pages 359--366, 2014.

\bibitem{Sagraloff201646}
Michael Sagraloff and Kurt Mehlhorn.
\newblock Computing real roots of real polynomials.
\newblock {\em Journal of Symbolic Computation}, 73:46 -- 86, 2016.

\bibitem{Yap:book}
C.K. Yap.
\newblock {\em Fundamental Problems of Algorithmic Algebra}.
\newblock Oxford University Press, 2000.

\bibitem{Ye1994271}
Yinyu Ye.
\newblock Combining {B}inary {S}earch and {N}ewton's {M}ethod to {C}ompute
  {R}eal {R}oots for a {C}lass of {R}eal {F}unctions.
\newblock {\em J. Complexity}, 10(3):271 -- 280, 1994.

\end{thebibliography}
}

\end{document}